\documentclass[sigconf]{acmart}

\setcopyright{none}
\renewcommand\footnotetextcopyrightpermission[1]{} % removes footnote with conference information in first column

\AtBeginDocument{%
  \providecommand\BibTeX{{%
    \normalfont B\kern-0.5em{\scshape i\kern-0.25em b}\kern-0.8em\TeX}}}

\usepackage{booktabs}
\usepackage{mathrsfs}
\usepackage{multirow}
\usepackage{subfig}
\usepackage{bm}
\usepackage{float}
\usepackage{color}
\usepackage{makecell}
\usepackage{listings}
\usepackage{caption}
\usepackage[linesnumbered,ruled,vlined]{algorithm2e}

\newcommand{\ShflBW}{\textit{Shfl-BW}}
\newcommand{\ShflBWfullname}{\textit{Shuffled Block-wise sparsity}}
\newcommand{\flexibility}{\textit{flexibility}
}
\newcommand{\performability}{\textit{computation efficiency}}

\usepackage[normalem]{ulem}
\newif\ifmodify
\ifmodify
\newcommand{\dl}[1]{\textcolor{red}{\sout{#1}}}

\newcommand{\f}[1]{\textcolor{orange}{#1}}

\newcommand{\guyue}[1]{\textcolor{brown}{#1}}
\newcommand{\comments}[1]{( {#1} )}
\else
\newcommand{\dl}[1]{}

\newcommand{\f}[1]{#1}

\newcommand{\guyue}[1]{#1}
\newcommand{\comments}[1]{\iffalse {#1} \fi}
\fi

\begin{document}

\title{Shfl-BW: Accelerating Deep Neural Network Inference with Tensor-Core Aware Weight Pruning
}
\author{Guyue Huang} \affiliation{University of California, Santa Barbara} \authornote{Email: guyue@ucsb.edu. This work was done when Guyue Huang was an intern at Alibaba DAMO Academy.}
\author{Haoran Li} \affiliation{Alibaba DAMO Academy}
\author{Minghai Qin} \affiliation{Western Digital Research} \authornote{This work was done when Minghai Qin was with Alibaba DAMO Academy.}
\author{Fei Sun} \affiliation{Alibaba DAMO Academy}
\author{Yufei Ding} \affiliation{University of California, Santa Barbara}
\author{Yuan Xie} \affiliation{Alibaba DAMO Academy}

% \author{Guyue Huang$^1$, Haoran Li$^2$, Minghai Qin$^{3}$}
% \authornote{Work done when Minghai Qin was with Alibaba DAMO Academy.}
% \author{Fei Sun$^2$, Yufei Ding$^1$, Yuan Xie$^1$}
% \author{$^1$University of California, Santa Barbara}
% \author{$^2$Alibaba DAMO Academy, $^3$Western Digital Research}
% \email{guyue@ucsb.edu}

\begin{abstract}
Weight pruning in deep neural networks (DNNs) can reduce storage and computation cost, but struggles to bring practical speedup to the model inference time.
\iffalse \comments{\f{(specific to irregular pruning?)} \guyue{no, not meant to be specific.}}\fi 
Tensor-cores can significantly boost the throughput of GPUs on dense computation, but exploiting tensor-cores for sparse DNNs is very challenging. 
% We identify two main reasons: 
Compared to {existing} CUDA-cores, tensor-cores require higher data reuse and matrix-shaped instruction granularity, both difficult to yield from sparse DNN kernels. 
Existing pruning approaches fail to balance the demands of accuracy and efficiency: random sparsity preserves the model quality well but prohibits tensor-core acceleration, while highly-structured block-wise sparsity can exploit tensor-cores but suffers from severe accuracy loss. 

In this work, we propose a novel sparse pattern, {\ShflBWfullname} ({\ShflBW}), designed to efficiently utilize tensor-cores while minimizing the constraints on the weight structure. Our insight is that row- and column-wise permutation provides abundant flexibility for the weight structure, while introduces negligible overheads using our GPU kernel designs. We optimize the GPU kernels for {\ShflBW} in linear and convolution layers. Evaluations show that our techniques can achieve the state-of-the-art speed-accuracy trade-offs on GPUs. For example, {with small accuracy loss}, we can accelerate the computation-intensive layers of Transformer~\cite{vaswani2017attention} by \textbf{1.81, 4.18 {and} 1.90$\times$} on NVIDIA V100, T4 and A100 GPUs {respectively} at 75\% sparsity.

% The fact that Deep Neural Networks can be greatly sparsified, despite its potentials in reducing model volume and accelerate inference, does not turn into actual speedup in real hardware due to the inherent low data reuse in sparse workloads. This performance gap is further enhanced by the tensor-core units in modern AI hardware, which improves FLops and puts pressure on the bandwidth and requires more data reuse. We find that existing pattern pruning methods fail to combine the need of algorithm and performance worlds, either allows too much irregularity that prohibits exploitation of TCUs, or puts too much constraints on the weight pattern to cater for TCU acceleration. 
% Weight pruning can reduce the model size and computation complexity, but fails to bring speedup on GPUs because sparsified networks cannot leverage lightning-fast tensor-cores. Non-structured pruning can preserve the model quality best but prohibits tensor-unit acceleration, while extremely-structured block-wise pruning can take advantage of tensor-cores but harms model quality severely. Our insight is that a loosely-structured pattern balances the quality and efficiency. We propose scattered-CSB, a novel sparsity pattern converted from the block-wise pattern with row- and column-wise reorder. We prune neural networks to this pattern based on progressive pruning and the Grow-and-Prune technique. We develop GPU kernels to exploit the implicit block-wise pattern and execute with tensor-cores. In terms of model quality, we achieve xxx. In terms of efficiency, we achieve xx.
\end{abstract}

\maketitle
\pagestyle{plain} % removes running headers

\section{Introduction} \label{sec:intro}
Deep neural networks (DNNs) have achieved state-of-the-art results in many domains including speech~\cite{amodei2016deep}, vision~\cite{szegedy2015going,he2016deep} and language~\cite{wu2016google,vaswani2017attention,brown2020language}. However, DNNs with superior results often have high computation and storage cost. OpenAI summarizes an exponential growth in the operation count and the number of parameters in the state-of-the-art DNN models~\cite{openaiblog}, and many giant language models have recently emerged~\cite{brown2020language,fedus2021switch}. Since even the most high-end computing platforms struggle to deploy those advanced models~\cite{shoeybi2019megatron}, their practical adoption is neither technically easy nor economically feasible. 

\begin{figure}[t]
    \includegraphics[width=\linewidth]{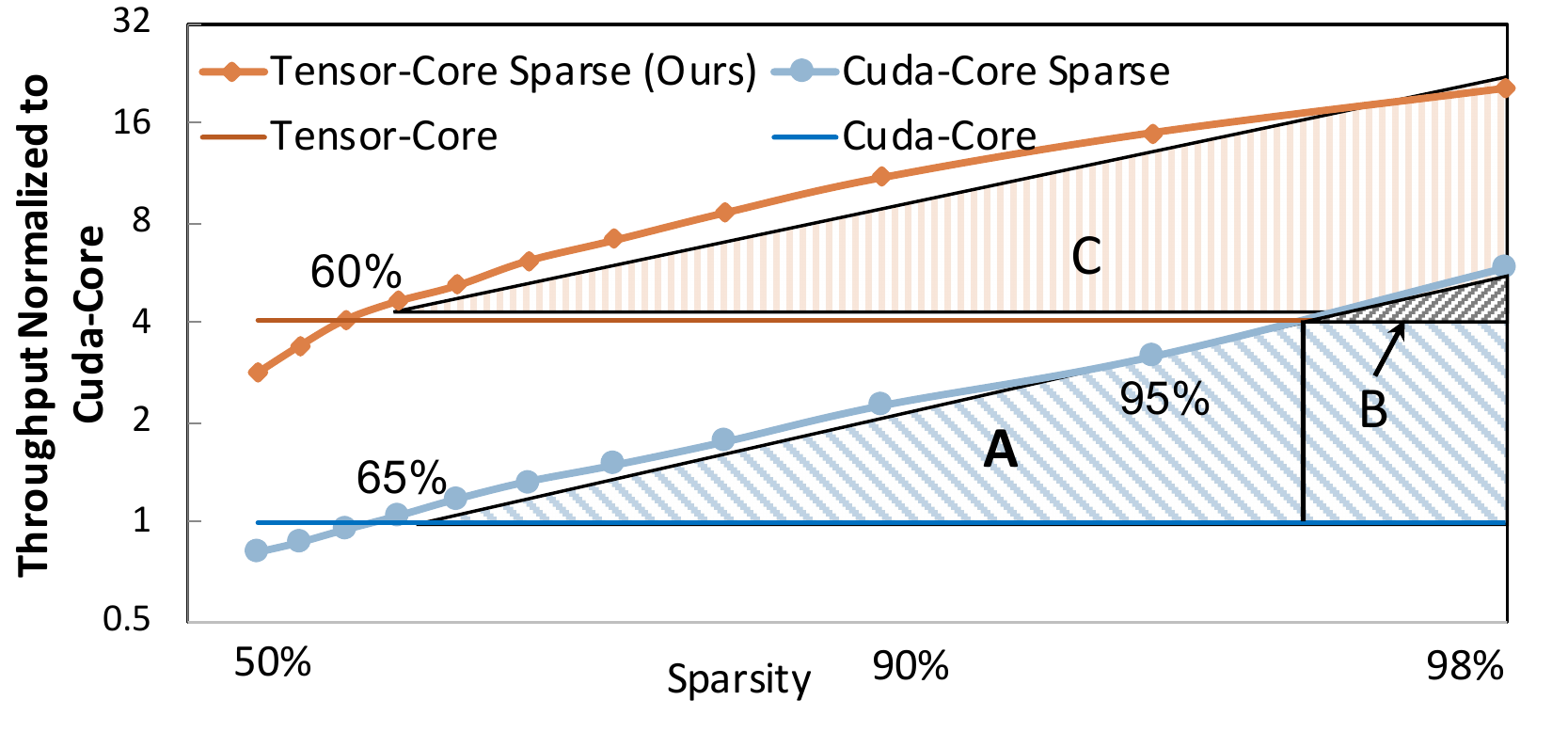}
    \caption{Throughput of Sparse matrix-matrix multiplication (SpMM) compared to dense GEMM. This paper provides the \textit{Tensor-Core Sparse} curve, which reduces the threshold where sparsity starts to show benefit. } \label{fig:1}
\end{figure}
\begin{figure}[t]
    \includegraphics[width=\linewidth]{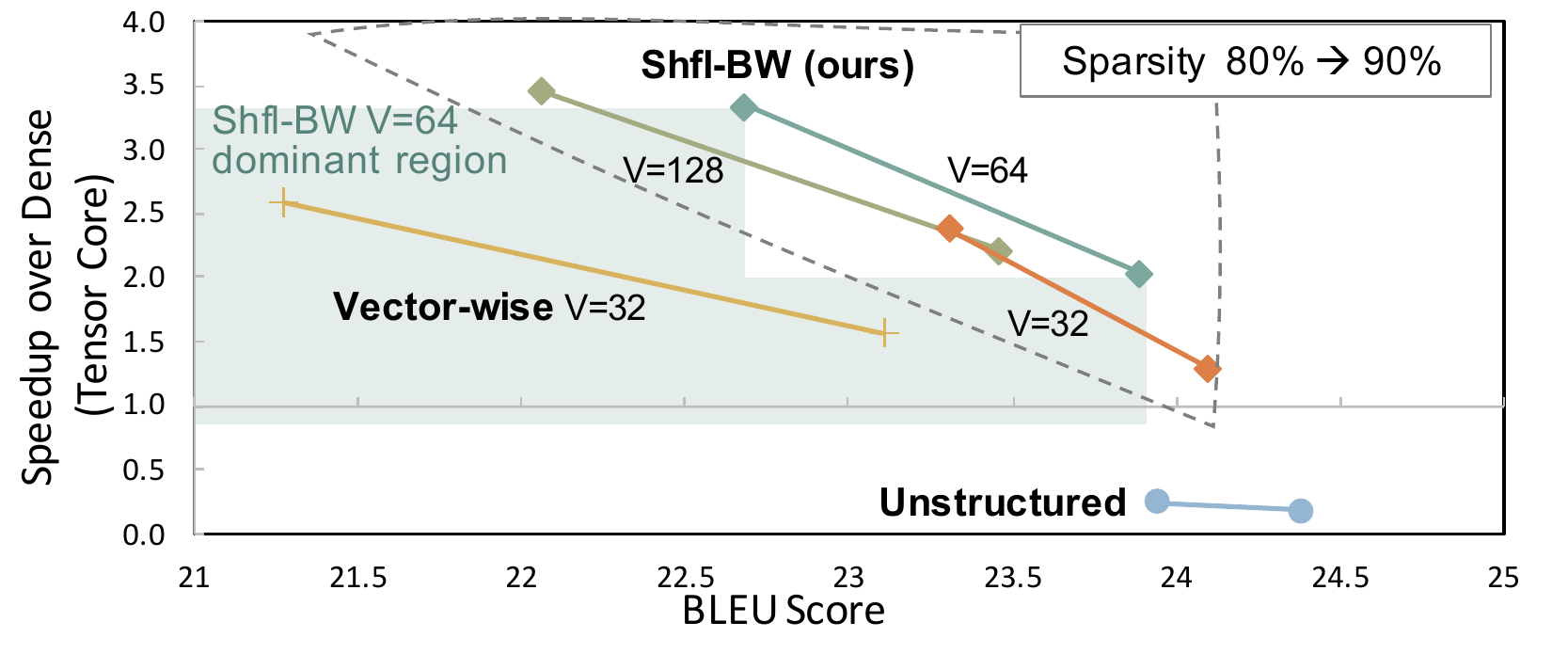}
    \caption{Accuracy-speedup trade-off curve of GNMT model\cite{wu2016google} on V100 GPU. Our proposal, {\ShflBW} can achieve practical speedup over the tensor-core based dense baseline while unstructured sparsity cannot. {\ShflBW} also provides a better trade-off compared to block-wise sparsity.} \label{fig:2}
\end{figure}

To turn the algorithmic advances into practical applications, researchers often exploit sparsity for efficient inference, targeting faster execution and less storage. Weight pruning~\cite{lecun1990optimal,han2015learning}, which removes unimportant connections in the neural networks, can effectively reduce the operation count and the model size. 
% Weight pruning is a promising approach to reduce model size and accelerate inference. Through removing unimportant neurons and connections in DNN models, the number of parameters and operations can be reduced simultaneously. Many existing papers study the inherent sparsity in DNN models and methods to induce them. It is often feasible to induce more than 70\% sparsity in state-of-the-art AI models and introduce little accuracy loss if combined with advanced training techniques. 

Reduced computation, however, does not necessarily lead to speedup. Sparse DNN kernels utilize \guyue{GPU} poorly, due to irregular memory access patterns and lower data reuse than its dense counterparts. \iffalse \dl{compared to dense DNN kernels}. \dl{With irregularity, fully utilizing the computation units at instruction-level is difficult, and with lower data reuse, data movement often becomes the bottleneck instead of computation.}\fi 
% They make it difficult to fully utilize the computation units. 
\guyue{On unstructured sparse matrices,} the vendor library cuSPARSE~\cite{cusparse} requires $>98\%$ sparsity to demonstrate practical speedup \guyue{over the dense baseline} on NVIDIA GPUs~\cite{Gale2020} \iffalse\f{(this is irregular? if so, need to mention)}\fi. \guyue{With optimizations tailored for} DNN's moderate sparsity, prior arts can only reduce this threshold to around $70\%$~\cite{Gale2020}. 

The tensor cores introduced in recent GPUs significantly boost their computation capabilities, making it more challenging to close the sparse-dense efficiency gap. Fig.~\ref{fig:1} shows \guyue{the throughput of SpMM, which is the kernel inside a weight-pruned linear layer. The shape of GEMM is $M/N/K=2048/128/2048$, and we use Sputnik~\cite{Gale2020} as the sparse CUDA-core implementation. If both computed with CUDA-cores, SpMM is faster than the dense baseline when the sparsity $>65\%$(region \texttt{A} in Fig.~\ref{fig:1}). However, \textit{CUDA-core based} SpMM is less performant than the \textit{tensor-core based} GEMM until reaching $>95\%$ sparsity (region \texttt{B}). Under such a high sparsity, undesired model accuracy loss is likely to prevent sparse models from being used.}
% \dl{compared to the dense DNN kernels running on tensor-cores, the advantageous region of sparse DNN without using the tensor-core shrinks to around $>95\%$ sparsity, \mh{in which undesired model accuracy loss~\cite{Guo2020} is likely to appear to prevent such sparse models from being used}.
% \dl{where people would hesitate to use due to the model accuracy loss}.} 
This clearly motivates a tensor-core-based sparse DNN solution to justify the performance benefits at a quality-acceptable sparsity-level , which is our focus in this paper(region \texttt{C}). 

% Closing the sparse-dense efficiency gap is much more challenging on tensor-cores than previously CUDA-cores. To identify these challenges,  rethink the irregularity and low data reuse under the tensor-core settings: tensor-core operates on coarse-grained \guyue{matrix-shaped operands instead of vector operands}\iffalse{\hr{blocks instead of} vector{s} or operands}\fi, and the irregular operations in sparse kernels struggle to efficiently fit into its granularity. Moreover, tensor-cores provide a much higher peak throughput that the only way to feed data fast enough is through enhancing data reuse. Unfortunately, sparse DNN kernels expose limited data reuse to exploit. \f{(this paragraph may be deleted if space is an issue)}

% Our solution is along the direction of {structured} pruning. Structured pruning {constrains} the structure of {the} pruned (or kept) weights to assist sparse kernel implementation, but unavoidably hurts model accuracy. 
\guyue{Unstructured sparsity cannot {take good advantages of} the tensor-core because it provides little data reuse opportunities and makes sparse kernels memory-bound. Although we can find GPU sparse kernels that manage to use tensor-cores, these kernels have strong requirements on the non-zero structure of sparse matrices, such as a block-wise structure~\cite{DeWit2012}. Such a highly-structured pattern, when used in DNN model pruning, involves high accuracy drop. Balanced sparsity~\cite{yao2019balanced,cao2019efficient}, recently supported by the NVIDIA A100 GPU~\cite{Nvidia2020}, {is limited by the fixed sparsity level (e.g. only 50\% on A100)}. Moreover, balanced sparsity still faces the memory-bound issue, since redundant data still need to be loaded from DRAM before effective operands are selected out. }
%  does not solve \guyue{the same data reuse challenge as unstructured sparsity, 
% 
% Previous methods fail to achieve strike this balance.  Block-wise sparsity {can} utilize tensor-cores {at} the cost of large accuracy loss \iffalse{\dl{compared to unstructured sparsity~\cite{Guo2020}}}\fi. 
% \comments{\f{(that is different from unstructured sparsity, which does not read redundant data, but cannot leverage SIMD.)}} 
In summary, towards the goal of tensor-core aware pattern pruning, prior sparsity patterns fail to balance {\flexibility} and {\performability}. \guyue{As we will analyze in Section~\ref{sec:format}, {\flexibility} can be estimated by the number of candidate structures, and {\performability} can be measured by operation intensity (FLOP-per-byte).} \iffalse the two critical metrics that we will detail in Section~\ref{sec:format}\fi 
% \comments{\mh{SINCE ``FLEXIBILITY'' AND ``PERFORMABILITY'' APPEAR MULTIPLE TIMES IN THE INTRODUCTION, WE SHOULD MENTION WHAT THEY ARE IN ONE OR TWO SENTENCES.}\guyue{todo}}

In this paper, we carefully design the sparsity pattern to simultaneously exploit tensor-core acceleration and minimize its constraints on the weight structure. We propose a novel sparsity pattern, {\ShflBWfullname} {({\ShflBW})}. \guyue{Fig.~\ref{fig:2} shows the advantages of {\ShflBW} over other sparsity patterns. Firstly, {\ShflBW} can achieve practical speedup over the tensor-core based dense baseline, while unstructured sparsity cannot. Secondly, {\ShflBW} dominates block-wise sparsity in {\performability}-{efficiency} trade-off.} %\f{(consider move Fig 2 to second page?)}

{\ShflBW} is a mutation of the block-wise sparsity with column- and row-wise permutation. The intuition behind {\ShflBW} is to preserve the {potential} data reuse in Block-wise sparsity, but simultaneously enhance the {\flexibility}. Our key insights are two-fold{s}: Firstly, the permutations we apply can asymptotically enlarge the {exploration} space of weight structure candidates. Secondly, with our GPU kernel designs, the permutation can be {performed} on-the-fly during the execution of sparse kernels with negligible {runtime} overheads. Specifically, we handle row-wise permutation with a reordered write-back phase at the end of the kernel, and handle column-wise permutation by stitching input in buffers to construct the same threadblock-scoped tiles as a dense kernel does. 

We practice {\ShflBW} sparsity on three DNN models, and demonstrate a better accuracy-speedup trade-off than existing pattern pruning methods. Given the recent trend {of adding} tensor-core-like units in processors to boost DNN workloads (AMD GPU~\cite{amd2020cdna}, Intel CPU~\cite{amx}), we expect our methodology and practice to have wider applications beyond NVIDIA GPUs. 

This paper makes the following contributions: 
\begin{itemize}
    \item We propose a novel sparsity pattern, {\ShflBWfullname} ({\ShflBW}), that balances {\flexibility} and {\performability} {towards tensor-core aware pattern pruning}. Our analysis shows that {\ShflBW} enlarges the number of candidate patterns over existing methods (unstructured, block-wise, balanced sparsity), and exposes the same opportunities of data reuse as block-wise and dense kernels. %has highest attainable throughput inferred from data reuse (same as Block-wise). 
    \item {On the} {inference speed} {side} , we {implement} {new} GPU kernels for the {\ShflBW} pattern to optimize sparse  {matrix multiplication and convolution} layers. Our techniques {transform} {\ShflBW} into Block-wise sparsity during execution on-the-fly with negligible {performance} overhead. 
    \item On the model accuracy side, we propose heuristic searching methods {to prune} weights into {\ShflBW} structures. {Experiments} show that {models pruned to} {\ShflBW} sparsity {achieves} less accuracy loss {than the} block-wise sparsity. 
    \item \guyue{Our GPU kernels can accelerate computation-intensive layers in DNN models by up to \textbf{1.81, 4.18, 1.90$\times$} on NVIDIA V100, T4 and A100 GPUs at 75\% sparsity, respectively. }Combined with pruning algorithm, we achieve a new state-of-the-art accuracy-speed trade-off for those  popular DNN models. 
\end{itemize}

\section{Background and Related Work}\label{sec:bg}

\subsection{Tensor-Core}
The latest NVIDIA Tensor-cores can perform a matrix-multiplication-addition (MMA) of granularity $M/N/K=16/8/16$ in one instruction. The peak throughput of tensor-cores exceeds original CUDA-cores by a large margin, e.g., $4\times$ on V100 and A100 GPU for half precision. Since tensor-cores can effectively accelerate matrix multiplication, they are included in the road-maps of many other processors targeting DNN workloads (e.g. AMD GPUs~\cite{amd2020cdna} and Intel CPUs~\cite{amx}). 

The most computation-intensive operations in many DNN models is General Matrix-Multiplication (GEMM), and when weights are pruned GEMM turns into an SpMM (Sparse-Dense Matrix Product). 
Tensor cores are very efficient on dense GEMMs, but it is difficult to leverage them in sparse computation, e.g. SpMM. We identify the main challenge is the {high demand on data reuse.} Since tensor-cores greatly boost the computation throughput but do not change the global bandwidth, it is crucial to increase data reuse, i.e. performing more computation on the loaded values. For example, given the A100 tensor-core throughput and last-level-cache bandwidth ~\cite{Nvidia2020}, one needs to perform $63$ MACs on each loaded value to achieve its peak throughput.
% In Figure.~\ref{fig:format}, we plot the tensor-core roofline based on A100 last-level cache bandwidth \comments{\f{(this plot is not informative)}}. It indicates that the reuse ratio should exceed 62 MAC (multiplication-addition) per loaded value to fully utilize tensor-cores \comments{\f{(where is it in the plot?)}}. 
\guyue{SpMM typically exposes much smaller data reuse opportunities, because each value in the dense input matrix is multiplied by fewer values in the other matrix due to sparsity.}

\subsection{Previous Work}
Prior efforts on efficient DNN inference focus on one or more aspects among algorithm, software and hardware. \textbf{Algorithm-only}: early studies explore the opportunities of weight pruning with one-shot pruning~\cite{lecun1990optimal,han1510compressing}, and recent studies improve pruning techniques via  
% some work explores how to sparsify DNN networks with one-shot~\cite{lecun1990optimal,gale2019state}, progressive methods~\cite{han1510compressing}, 
% or at early stage of training~\cite{frankle2018lottery,you2019drawing}. Recent studies also explore 
regularization~\cite{zhang2018systematic} and weight pattern exploration~\cite{ma2021effective}. This thread of work usually focus on theoretically reducing effective operations in DNN. However, unstructured sparsity can be hard to transform into practical speedup (Figure~\ref{fig:1}). \textbf{Software-only}~\cite{Gale2020} and \textbf{hardware-only}~\cite{parashar2017scnn,han2017ese,hegde2019extensor} artifacts focus purely on accelerating unstructured sparse kernels in DNN. % Since they do not cooperate with algorithm design, they must only assume \textbf{unstructured sparsity}, i.e. arbitrary non-zero patterns.
Instead of trying to accelerate unstructured sparse kernels, the \textbf{algorithm-hardware co-design} approaches propose specialized sparsity patterns that are hardware-friendly, on accelerators~\cite{cao2019efficient} or GPUs~\cite{yao2019balanced,Nvidia2020}. \textbf{Algorithm-software co-design} approaches coordinate pruning pattern and kernel optimizations on GPUs~\cite{Guo2020} or CPUs~\cite{elsen2020fast}. 

\begin{figure}[t]
    \includegraphics[width=\linewidth]{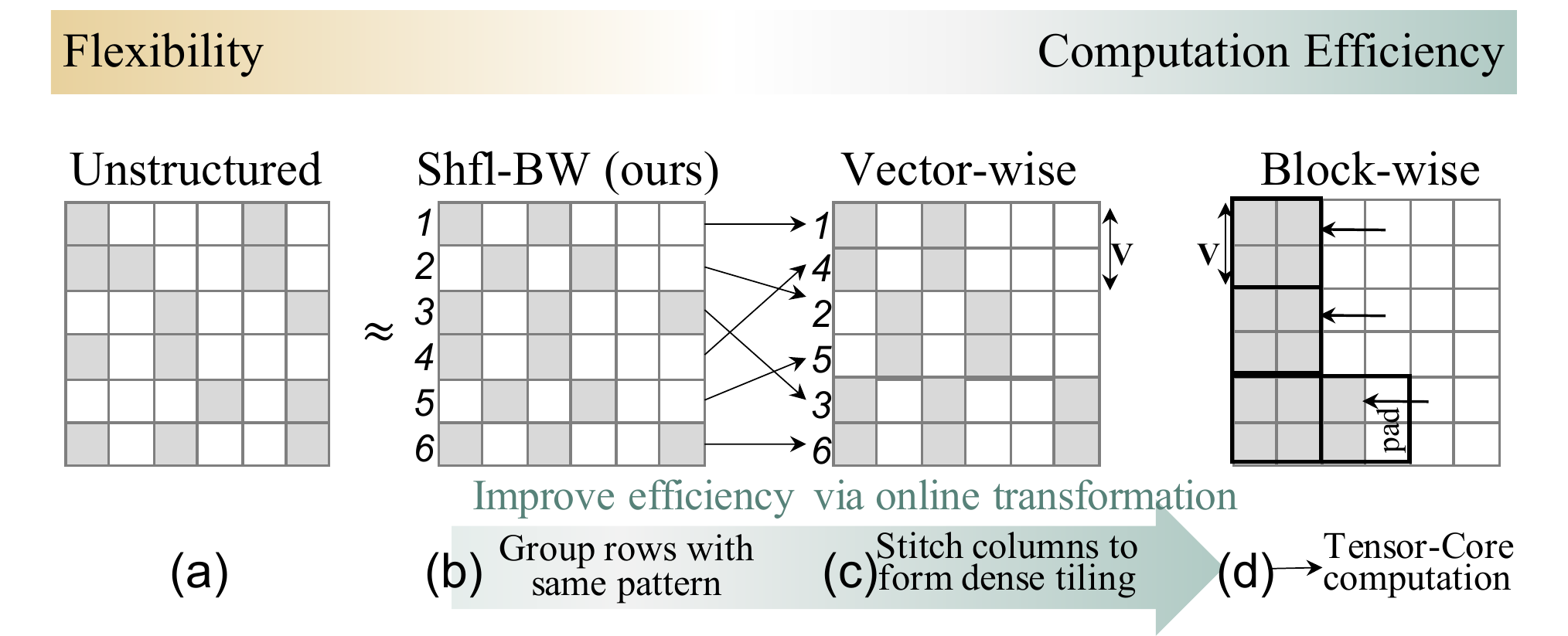}
    \caption{Different sparse patterns and how to transform from {\ShflBW} to block-wise sparsity.} \label{fig:format}
\end{figure}

Previous studies have proposed many sparsity patterns. \textbf{Block-wise} sparsity requires non-zero weights to form block shapes. An example is the sparse matrix in Figure~\ref{fig:format}(d), where an entire block of $V\times V$ parameters is either kept or pruned together. An extension of the block-wise sparsity is \textbf{vector-wise} sparsity (Figure~\ref{fig:format}(c)), i.e. pruning granularity is a $V \times 1$ vector. \textbf{Balanced} sparsity allows $m$ non-zeros within a local shape of $n$ values. For example, the NVIDIA A100 GPU optimizes its tensor-core for the 2-in-4 balanced sparsity: 2 out of consecutive 4 elements in one row are non-zero.

% This paper also uses structured pruning, specifically targeting tensor-core-enhanced DNN processors like NVIDIA GPUs. Our proposed sparsity pattern, {\ShflBW}, is an extension of the block-wise sparsity. The most-related prior work is the Tile-Wise sparsity~\cite{Guo2020}, which combines vector-wise with coarse-grained neuron pruning and unstructured sparsity, but does not consider row-wise shuffling as we do. We have provided comparison to this work in Section.~\ref{sec:eval}.

% \subsection{Pattern Pruning}
% In the context of weight pruning, \textbf{unstructured} sparsity means not setting any constraint on the location of non-zero weights, as shown in Figure~\ref{fig:format}(1). To exploit the reduced computation for real speedup in hardware, structured or pattern pruning is proposed. \textbf{Block-wise} sparsity requires non-zero weights to form block shapes. As shown in Figure~\ref{fig:format}(4), an entire block of $V\times V$ parameters is either kept or pruned together. An extension of the block-wise sparsity is vector-wise sparsity, i.e. pruning granularity is a $V \times 1$ vector. %Block-wise sparsity exposes high data reuse opportunities since the  . However, the highly-structured pattern  on the patterns of non-zeros. 
% \textbf{Balanced} sparsity allows $m$ non-zeros within a local shape of $n$ values. For example, the NVIDIA A100 GPU optimizes its tensor-core for the 2-in-4 balanced sparsity: 2 out of consecutive 4 elements in one row are non-zero.

\section{The {\ShflBW} Sparse Pattern} \label{sec:format}

\subsection{{\ShflBW} {Pattern}}
The key problem in this paper is designing a sparse pattern that balances algorithm and computation need. We need one pattern to efficiently utilize tensor-cores, and at the same time, flexible enough to mimic unstructured sparsity. In existing sparse patterns, these two aspects are hard to coordinate. Figure~\ref{fig:format} shows three existing patterns and {\ShflBW}, in the descending order of flexibility. The left-most is unstructured sparsity, which sets no pattern constraints at all but are least computation-friendly. Going right, the patterns add more constraints to the weight structure, but the computation efficiency is also improved. Block-wise sparsity is the most computation-friendly, because we can tile it to dense sub-matrices and apply tensor-core based dense kernels on each tile. However, it requires non-zero weights to cluster into blocks and can lead to severe accuracy loss. 

Our proposal is {\ShflBWfullname} ({\ShflBW}). It enjoys the flexibility of row- and column-wise permutations on top of block-wise sparsity. We manage to recover the computation efficiency through transforming {\ShflBW} into block-wise \textit{online during kernel execution}. Fig.~\ref{fig:format}{~(b)} shows an example of {\ShflBW} in matrix. At the first look, non-zeros in matrix {\texttt{(b)}} appear in random positions  in {the} matrix, looking like unstructured sparsity, but they are actually organized by the following rules: through row-wise permutation, we can group together rows with the same non-zero positions, and get a vector-wise sparse matrix with ($V\times 1$) vector shape like matrix {(as shown in Figure~\ref{fig:format}~(c))}. Next, through column stitching within each group of $V$ rows, we can get a block-wise sparse matrix with $V \times V$ block shape like matrix {(Figure~\ref{fig:format}~(d))}. These two steps  not only {explain} {\ShflBW}'s definition, but also are our methods to transform {the matrix format from} {\ShflBW} to block-wise in an SpMM kernel. 

% The \textit{Shfl-BW} pattern is transformed from block-wise pattern via column- and row-permutations. Figure~\ref{fig:format} shows a block-wise sparse matrix with $V\times V$ blocks. Firstly, we treat each $V$-row part as a sub-matrix, and {shuffle} the columns within the sub-matrices. This gives the vector-wise sparse pattern in Figure~\ref{fig:format}\comments{ \f{(you need to give a label and reference here)}\guyue{ todo}}. After column shuffling, we apply a row shuffling to the entire weight matrix. This gives the \textit{Shfl-BW} pattern as shown in Figure~\ref{fig:format}. 

% The key problem that this paper tackles is to design a sparse pattern that efficiently utilizes tensor-cores, while reducing the accuracy loss caused by weight structure constraints. In the rest of this section, we analyze the advantage of {\ShflBW} over other sparse patterns from two aspects: \textbf{flexibility} and \textbf{performability}. \comments{\f{(this can be removed)}}

\subsection{Analysis}

\subsubsection{Flexibility} \label{sec:format:flexibility}

To improve the model accuracy under a fixed sparsity, the weight structure should be {flexible} enough to cover more important weights. 
% We do not evaluate the flexibility based on the accuracy of  {the pruned} models, because it is hard to decouple the effect of sparse pattern and training techniques in the final accuracy. To evaluate the quality of sparse pattern a priori, 
We use the number of candidate structures under a certain sparsity to quantify \textit{flexibility} of a sparse pattern. 
% Suppose the sparsity is $(1-\alpha)\in (0,1)$, i.e. non-zeros taking up $\alpha$ of the matrix. Let $F(\alpha)$ denote the number of candidate structures.  %The metric is defined as the number of valid weight structures under a fixed sparsity. 
% % This metric is based on the intuition that, assuming there exist advantageous sub-networks in the original dense network that can be trained to high accuracy ~\cite{frankle2018lottery}, a sparse pattern with more candidate structures has a higher probability to contain such advantageous sub-networks.
% The number of candidate structures in block-wise (or vector-wise) sparse matrix equals to the combination of $V\times V$ (or $V\times 1$) sub-matrices.
% % \comments{\f{(this is the first time density appears, should define it, also, mention without loosing generality, M, K should be divisible to V)}}. 
% Hence, $F_{blockwise}(\alpha)$ is the number of combination $C_{N_{block}}^{\alpha {N_{block}}}$ where ${N_{block}} = \frac{M}{V} \cdot \frac{K}{V} $ \f{refers to}  all possible block positions. \f{Follow\hr{ing} the same method, we calculate} the candidate space size of vector-wise sparsity is $C_{N_{vector}}^{\alpha {N_{vector}}}, \hr{where} N_{vector}=\frac{M}{V}\cdot K $. 
{\ShflBW} performs row shuffling on top of vector-wise sparsity. For the weight matrix of $M$ rows, 
% the row-permutation has $M!$ candidates, but permutation within the $V$ rows of the same pattern yields same structures. 
% Hence the candidate space size of {\ShflBW} is 
% $$|candidate\_set|_{{\ShflBW}, \alpha}= |candidate\_set|_{{vectorwise}, \alpha} \times \frac{M!}{(V!)^{\frac{M}{V}}}$$. 
the candidate space is multiplied by $\frac{M!}{(V!)^{\frac{M}{V}}}$, which is the number of combinations which partition $M$ rows into groups of size $V$. 
Row-wise shuffling significantly enlarges the candidate space. For example, for a weight matrix with $M=512$ rows and when $V=128$, this combination number already exceeds $e^{700}$. 
\comments{\f{(this is difficult to understand with such short description)}\guyue{todo}}

\subsubsection{Computation Efficiency}\label{sec:format:perform}
To increase the utilization of tensor-cores, the weight structure {needs to} {expose} the opportunities of data reuse. We evaluate the \textit{computation efficiency} of a sparse pattern by operation intensity, which equals to dividing the total number of float-point operations by number of bytes loaded from GPU global memory. The data reuse is {achieved} through tiling. In an SpMM kernel, suppose a GPU threadblock first loads a $T_M\times T_K$ tile of the left-hand input matrix, and $T_K \times T_N$ from the right-hand matrix, to the on-chip shared memory. Next, the threadblock multiplies the loaded tiles and accumulate on a $T_M\times T_N$ output tile stored in the register file. We assume the maximum tiling size as long as the register file is large enough to hold the output accumulators. 

We discuss two situations: when the tiled sparse matrix is dense or still sparse. Under unstructured or balanced sparsity, since the non-zeros {values are} distributed randomly and not forming any dense structures, the tiled sparse matrix is still sparse. \guyue{We make the tiling size maximum subject to the condition that the register file is large enough to hold the output accumulators.}
% We assume \dl{the maximum tiling size as long as} the register file is large enough to hold the output\f{s for the largest tiling size}\dl{ accumulators}.
% Figure~\ref{fig:reuse} shows the differences between unstructured/balanced SpMM and block-wise/{\ShflBW} SpMM. To {increase} data reuse, we buffer a tile of weights and activations in the GPU shared memory. If the weight matrix is unstructured- or balanced-sparse, a weight tile of $T_M \times T_K$ \comments{\f{(what is $T_M$? it is not described)}\guyue{todo}} only contains a proportion of $\alpha$ non-zeros, $\alpha$ being the density \hr{(may consider use (1-sparsity) to replace density everywhere to avoid confusion)}.
To {find out the} maximum data reuse is {equivalent} to solve the following problem:
$${Max\_reuse = }\max_{T_M,T_N}(\frac{2\alpha T_M \times T_N \times T_K\text{Flop}}{\alpha (T_M\cdot T_K) + (T_K \cdot T_N)\cdot 2\text{Byte}}),$$
subject to$$ {T_M \times T_N \leq Size\_regfile}$$
\guyue{where $\alpha \in (0,1)$ denotes the non-zero ratio.}
Solving this optimization problem gives us $Max\_reuse = \sqrt{\alpha} Reuse_{dense}$ where  
% \dl{$T_{opt,dense} = \sqrt{Size\_regfile}$,} 
$Reuse_{dense} = \frac{T_{opt,dense}}{2} flop/byte$ is the maximum reuse in a dense GEMM{, and $T_{opt,dense} = \sqrt{Size\_regfile}$}.
% $$Max\_reuse = \sqrt{\alpha} Reuse_{dense}, T_M = \frac{1}{\alpha}T_N = \frac{1}{\sqrt{\alpha}} T_{opt,dense}$$ where $T_{opt,dense} = \sqrt{Size\_regfile}$ is the optimal square-tiling size in dense GEMM and $Reuse_{dense} = \frac{T_{opt,dense}}{2} flop/byte$. 

On the other hand, under block-wise, vector-wise and our {\ShflBW} sparsity, the tiled sparse matrix can be dense. Block-wise matrix is dense within $V\times V$ non-zero blocks, thus if {we} make $V\times V$ the tiling size, the tiled sparse matrix is dense. The data reuse ratio can reach $Reuse_{dense}$ as long as $V \geq T_{opt,dense}$. Vector-wise and {\ShflBW} matrices can be transformed online to block-wise, and achieve the same data reuse ratio. %If the weight matrix is in {\ShflBW} {format}, through the online stitching technique we will introduce \hr{in} (Section~\ref{sec:kernel}), the threadblock tile looks the same as block-wise SpMM, hence the data reuse is the same. The actual reuse ratio is $\frac{T_{M}}{2} flop/byte$, \dl{which is} the same as in a dense GEMM {computation} with same tiling shape if $T_M \geq T_{opt,dense}$, which is achievable in our experiments.

In summary, in terms of computation efficiency, {\ShflBW}, vector-wise and block-wise {format} exposes higher data reuse opportunities then unstructured and balanced sparsity, because we can tile them into dense sub-matrices. In terms of flexibility, {\ShflBW} significantly  
% \hr{(what is the meaning of asymptotically here?)} 
enlarges the candidate space of block-wise and vector-wise sparsity. 

\section{GPU Sparse Kernel} \label{sec:kernel}
\begin{figure}[t]
    \includegraphics[width=\linewidth]{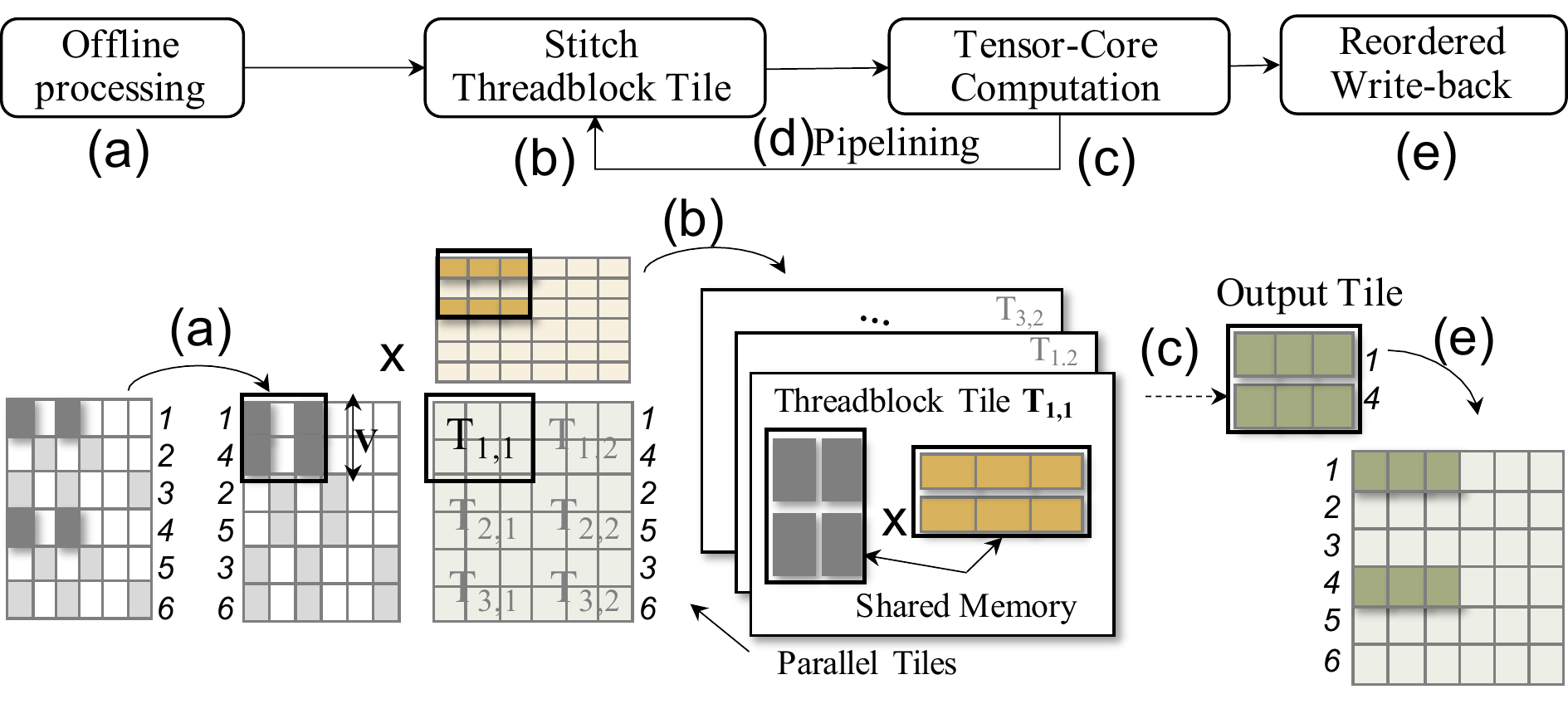}
    \caption{GPU kernel design of {\ShflBW} SpMM.} \label{fig:kernel}
\end{figure}

% \begin{figure}[h]
% \centering
%     \includegraphics[width=\linewidth]{fig/kernel.s1}
%     \caption{Steps in dense GEMM and {\ShflBW} SpMM kernels.\f{left is (a), right is (b), letter too small}} \label{fig:steps}
% \end{figure}

% \begin{figure}[h]
%     \centering
%     \includegraphics[width=\linewidth]{fig/kernel.s2}
%     \caption{Visualizing dense GEMM steps.}
% \end{figure}

% \begin{figure}[h]{}
%     \centering
%     \includegraphics[width=\linewidth]{fig/kernel.s4}
%     \caption{Reordered write-back to handle row-wise shuffling. \f{letter too small, have (a), (b), (c) to indicate different steps}}\label{fig:write-back}
% \end{figure}

% \begin{figure}[h]{}
%     \centering
%     \includegraphics[width=0.8\linewidth]{fig/kernel.s3}
%     \caption{Input stitching to handle column-wise shuffling.}
%     \label{fig:stitch}
% \end{figure}

% \begin{figure}[h]{}
%     \centering
%     \includegraphics[width=\linewidth]{fig/kernel.s5}
%     \caption{Pipeline optimization of {\ShflBW} SpMM. \f{legend, add (a), (b)... }}
%     \label{fig:pipeline}
% \end{figure}
\subsection{Overview}
Figure~\ref{fig:kernel} shows the key steps in a {\ShflBW} GPU SpMM kernel. Step \texttt{(a)} is an offline processing that transforms a {\ShflBW} matrix into {a} vector-wise sparse matrix and an array to store the original row-indices. Storing the sparse matrix in vector-wise format allows us to load values contiguously from the sparse matrix during kernel {execution}, benefiting the bandwidth efficiency. Step \texttt{(b)} load{s} tiles of both input matrices into the GPU shared memory. In this step, we perform in-buffer stitching, to be detailed in Section~\ref{sec:kernel:stitching}. \guyue{Next, in Step \texttt{(c)} we perform matrix-multiplication-addition (MMA) {using} tensor-cores.\guyue{ Since the loaded tiles are dense, this step is the same as it is in a dense GEMM. }We loop over Step \texttt{(b)} and \texttt{(c)} with our pipelining design (marked by \texttt{(d)}, detailed in Section~\ref{sec:kernel:prefetching}). We point out this kernel requires two-level pre-fetching for sparse metadata and the actual input data, complicating the pipeline schedule. }
%to consume all non-zero vectors in this sparse matrix lane. 
Finally, the output tile is written back to the global memory in Step \texttt{(e)}. Because we reorder the sparse matrix rows offline, in this step we perform online reordering and directly write results to the correct row{s} in the output matrix, to be detailed in Section.~\ref{sec:kernel:reorder}. 

The following sections detail our three designs using SpMM as example. We also implement 2D convolutions with similar designs using the \textit{implicit-GEMM} algorithm~\cite{cudnn}, where the input feature map is unfolded into a matrix form temporally in on-chip buffers.

\subsection{Reordered Write-Back}\label{sec:kernel:reorder}
In Figure~\ref{fig:kernel}, step \texttt{(a)} and \texttt{(d)} together \guyue{perform} the reordered write-back technique. 
%\comments{\f{(somewhat confusing, what is reordered write-back? the later text discusses it? why only two steps?)}}
In {\ShflBW} sparse matrix, rows with identical non-zero patterns are scattered to discontiguous positions. For example, in the figure, row $1$ and $4$ have identical patterns. If we compress and store this sparse matrix row-wise, and directly load from these two rows, this uncontiguous memory access pattern cannot fully utilize the global bandwidth. 
% In order to form dense tiles in the shared memory \f{(do we form dense tiles in the shared memory? we don't, right?)}, we \dl{need to} load weights from rows that have non-zero weights in the same column. This means we need to revert the row shuffling, and tile the input based on the original row order.  \f{( it is confusing, consider rewrite)} Figure.~\ref{fig:write-back} shows our reordered write-back technique. We do not tile the output matrix based on the shuffled row order, because we need the weight rows for each tile to have column-aligned non-zeros for data reuse. \f{(consider rewrite)} 
Instead, we \guyue{reorder {\ShflBW} to a vector-wise matrix store weights in the same vector contiguously.} 
% and store the weight matrix in the vector-wise form, 
We keep the original row-indices in an array \guyue{for reversing the row order in the output matrix}.
%\comments{\f{(this is confusing, what vector-wise form? you mean after performing row shuffle offline? need to be explicit)}}
Steps \texttt{(b,c,d)} perform {the} SpMM {calculation} assuming the input is a vector-wise sparse matrix 
\comments{\f{(why three steps perform the computation?)}}
. Finally, when writing the output back to global memory, we refer to the row-indices to directly write in the original row ordering in step \texttt{(e)}.
% When writing results to the output matrix in the last phase of the kernel, we refer to the shuffling indices to write to the correct row in the reordered final output. Figure~\ref{fig:write-back} shows an example: Multiplying a row-shuffled weight matrix is equivalent to multiplying the original weight and then row-shuffle the output \f{(this does not talk about figure 7)}. When implementing the reordered write-back, 
We exploit shared memory to buffer the shuffle indices for this tile, so that the number of global memory loads is reduced. %This process is denoted as the $ReorderedWriteBack$ in Algorithm~\ref{alg:spmm}.
% That being said, a naive way to first use unshuffled matrix to compute GEMM, and then use another kernel to reorder the output rows according to an index array. We optimize this by fusing the reorder step into the GEMM kernel, which requires us to locate the correct index while we do write back. In implementation, the row shuffle indices is passed as an argument to the kernel. Then it uses the index to compute the actual offset of the tile in the output matrix, which might be scattered to different rows. Finally, the write-back is done with this offset. In experiments, we find adding this step only adds <3\% overhead for all real model matrices we test.

\subsection{In-Buffer stitching}\label{sec:kernel:stitching}
Step \texttt{(b)} in Figure~\ref{fig:kernel} illustrates the in-buffer stitching technique. After the offline processing in step \texttt{(a)}, the sparse matrix is in vector-wise form. However, tensor-core requires that the reduction dimension in one MMA instruction is at least 8. In order to map to tensor-core granularity, we stitch multiple sparse vectors together. In Figure~\ref{fig:kernel}, we stitch two vectors in the sparse matrix to form a $2\times 2$ dense tile \guyue{just as an example, but in practice the chosen tile size is often $V\times16$ or larger.} When loading the dense matrix, we also stitch data from two discontiguous rows to form a tile. The stitching of the sparse matrix tile is done offline during compression, while the stitching of the dense matrix is done on-the-fly during the loading.

% The column reorder step transforms block-wise sparsity to vector-wise, and the distributed weight vectors need to multiply discontiguous rows in the input activation matrix, as shown in Figure~\ref{fig:stitch} \f{(the other way round?)}. Since the input activation is online generated in inference, statically condensing these rows into a new dense matrix and then invoking a normal GEMM has prohibitive cost. Moreover, the column indices of \f{the} non-zero vectors differ among \f{the} row tiles, hence the relevant rows in input activations are also different for different tiles. We call our approach in-buffer stitching. When loading weights, refer to the column indices to load from the right row of input activations. For example, when \f{the} threads in a threadblock collectively load a tile of values, one thread who is responsible to load \textit{into} the first row of the buffer tile, should load \textit{from} the row pointed by the first column index of weight in this tile. The discontiguous operand rows in input activation is \textit{stitched} in the shared memory, and after stitching we can invoke a normal $WarpMMA$ just like the tiles are loaded from dense input matrices. \f{(it is best to describe it using an example, Figure 8 is not discussed)}

\textbf{Discussion on the input activation layout:} \guyue{We assume the dense matrices in SpMM has row-major layout, because the memory read/write pattern to the input/output dense matrix are both row-wise. }
% Since the access pattern to the dense matrix is row-wise, storing the dense matrix in column-major layout causes severe performance drop due to discontiguous memory access. The reordered write-back mechanism also requires row-major layout of the output matrix. Hence we choose row-major layout for the dense input and output matrices. 
In real models, this means we make batch the innermost dimension of layout in linear layer and 2D convolution. This does not involve transposition overhead in convolution models \guyue{because convolutions often follow the previous layer's BatchNorm and is succeeded by another BatchNorm}. In models which apply LayerNorm and requires feature to be stored contiguously, transposition is necessary, but transposition 
% such as Transformer, the feature innermost layout of activations needs to be transposed to batch innermost before invoking our sparse linear kernels, but this operation 
can be easily fused into previous LayerNorm and involves negligible overhead~\cite{Guo2020}. %In \f{the} evaluation\f{,} \dl{we provide speedup assuming} \f{the comparison on the performance speedup assumes} both fused transposition (no overhead) and adding explicit transposition. \f{(you didn't mention adding explicit transposition before, where is it?)}

\begin{algorithm}[t]

    \SetAlgoLined
    \KwData{A.row\_indices[], A.metadata[], A.values[], B[K][N]}
    \KwResult{C[M][N]}
    
    metaload\_step = 0\;
    load\_step = metaload\_step-MetaPrefetchStage\;
    step = load\_step - (PipeStage+1)\;
    total\_step = nnz\_this\_lane/$T_K$\;
    \While{step < total\_step} {
        \If{metaload\_step \% MetaPrefetchStage == 0} {
            BulkLoadMeta(A.metadata[], metaload\_step)\;
        }
        \If{0 $\leq$ load\_step $\leq$ total\_step} {
            StitchTile(buffer[load\_step \% PipeStage])\;
        }
        \If{step $\geq$ 0} {
            WarpMMA(buffer[step \% PipeStage])\;
        }
    step++, load\_step++, metaload\_step++\;
    }
    ReorderedWriteBack(C, buffer, A.row\_indices[])\;
    \caption{GPU SpMM with {\ShflBW} sparsity pattern }
    \label{alg:spmm}
\end{algorithm}
\subsection{Prefetching Meta-Data}\label{sec:kernel:prefetching}
Pipelining is an important technique to hide memory loading latency. 
% In an original dense GEMM kernel, the latency of loading input tiles is \dl{hided} \f{hidden} through pipelining the main loop in Figure.~\ref{fig:steps}. 
In SpMM, we not only need to prefetch input tiles (as shown in Line 10 of Algorithm~\ref{alg:spmm}), but also the metadata of {the} sparse weight matrix, {which}, in our case, {is} the column indices of {the} non-zero weights. If we load {the} metadata and data of the same tile together, 
% \dl{{unhidden} data dependencies will still exist} 
{it creates an explicit data dependency between the metadata and the activation matrix}. This is because during in-buffer stitching, we refer to the column indices to load the relevant rows in the input activation and form the dense matrix tile. 
% \dl{If metadata is loaded with data, loading the dense matrix tile will have to wait until metadata arrives.} 
To resolve this dependency, we propose to prefetch metadata, i.e. loading metadata of future weight tiles ahead of time. 
% As shown in Figure~\ref{fig:pipeline}, in an unpipelined implementation, loading metadata, loading data (both weight and input) and computation are completely serialized due to data dependencies. If we only pipeline the inner loop in Figure~\ref{fig:steps} which contains data loading and computation, we get the second lane in Figure~\ref{fig:pipeline} , where the computation and data loading is overlapped, but the metadata and data loading are still serialized. 
Moreover, metadata usually has a small volume, and aggregating metadata of multiple tiles together leads to more efficient usage of bandwidth. Algorithm~\ref{alg:spmm} shows the pipelining strategy, where metadata is pre-fetched in bulks of $MetaPrefetchStage$ steps (Line 6-8). 

% % Our solution is the bottom lane, which uses both bulk loading and prefetch of the metadata. %This procedure is described in Algorithm~\ref{alg:spmm}, where $metaload\_step$, $load\_step$ and $step$(computation) are all ahead of the next one.

% \subsection{Implementing Convolution with Implicit-GEMM}

% conv is implemented using implicit gemm. To ensure loading alignment, we need batch first layout. In most convnets, batch first is used because it is beneficial to batchnorm, as the case in our resnet50 example. If not, we can still use the previous way, fusing the previous transpose in previous kernel. 

\section{Pruning Algorithm} \label{sec:pruning}
\begin{figure}[t]
    \includegraphics[width=\linewidth]{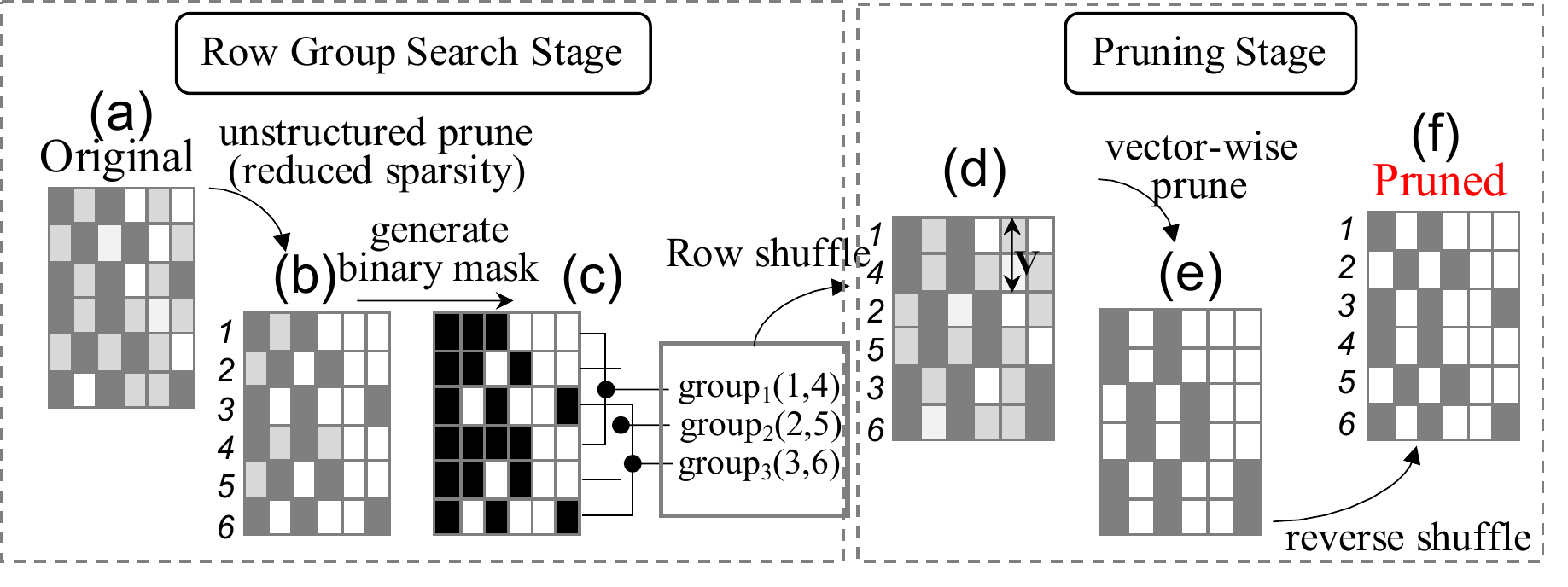}
    \caption{The {\ShflBW} pattern searching algorithm.} \label{fig:pruning_alg}
\end{figure}
This section introduces our pruning algorithm for {\ShflBW} sparse pattern. Given the importance scores of all weights, our algorithm decides which weights to keep following the {\ShflBW} pattern.The {\ShflBW} pattern introduces unique challenges to pattern searching due to the enlarged candidate space from row-wise shuffling. 
% This section introduces our method to prune a dense model into {\ShflBW} sparsity {pattern}. We call such a procedure as pattern search. In addition, a pruning workflow also include fine-tuning steps or other training techniques, and we follow existing work and do not introduce in this part. %\f{(this is optional, may remove to save space)} Algorithm~\ref{alg:pruning} describes the overall procedure of pattern searching. 
% Before {elaborating} Algorithm~\ref{alg:pruning}, we first describe the unique challenges of searching {\ShflBW} pattern, compared to searching block- or vector-wise pattern. 
% {\ShflBW} introduces unique challenges to the design of the pattern searching algorithm. 
In block-wise pattern searching, finding the combination of weight blocks with highest score can be simply done by a greedy method, i.e. selecting blocks with highest scores until the density is reached. However, {a greedy method does not work for the} {\ShflBW} {pattern}, because selecting a \textit{shuffled} block {implicitly enforcing} all {the selected} weights in these rows {to be} column-aligned. Formally stated, on an $M \times K$ score matrix 
% \f{(what is score matrix?)}
, our target is to find a row permutation $\pi$, such that when we shuffle all $M$ rows with $\pi$ and then apply vector-wise pruning, we can obtain {the} highest total scores. %This problem turns out to be NP-hard 
% \f{(do we have proof that this is NP-hard?)}
{Thus, w}e design heuristic methods to tackle this search problem. %\f{(is column shuffle only for one row?)}
% \f{(I'd like to rewrite this and previous paragraph)}

% In this section, we describe how to prune a dense model to \textit{Shfl-BW} sparse. Following any pruning approach, we want to prune weights with small absolute value. In pattern pruning, since often a group of weights are decided together whether to keep or drop, pruning is based on the score of the entire pattern entity instead of individual weights, often by summing-up the scores of each weights inside. 

% \textit{Shfl-BW}, however, stems out from all previous pattern pruning approach in the difficulty of deciding pruning patterns. While other pruning patterns can be easily divided in the weight matrix, for \textit{Shfl-BW} since we allow row shuffling, candidate structures to be pruned overlap with each other, and it is difficult to make globally optimal choice. Specifically, the problem can be formulated as 

% This problem is NP-hard. We thus propose a heuristic way to tackle it. Our solution is in Algorithm 1 and shown in Figure .

Figure~\ref{fig:pruning_alg} shows our two-step approach for searching {\ShflBW} pattern{: firstly, deciding the row shuffling, and secondly, apply{ing} vector-wise pruning to the shuffle matrix} 
%\f{(consider rewrite, do we shuffle columns?)}
. The question then becomes how to decide the row shuffling before we are aware of the vector sparse pattern. Observe that after row shuffling, the rows in the same row group will have the same vector column pattern. In other words, they keep the weights in the same columns. The heuristic here is that rows with weights appearing in similar column-wise {positions} should be reordered together, so that in the next stage, vector-wise pruning will keep more important weights. 
% \f{(consider rewrite, it is difficult to follow)}

The example in Figure~\ref{fig:pruning_alg} elaborates this algorithm. %A natural way is to cluster the rows according to the positions where important weights appear. We can either directly cluster the contiguous valued vectors, or 
The input to the algorithm is an importance score matrix, as shown in matrix \texttt{(a)}. We use the absolute value of each weight as its important score~\cite{han2015learning}. At the first step, we apply unstructured pruning to this score matrix, and keep the $\beta$ weights with highest importance scores. The target non-zero ratio in this step, $\beta$, can be different from the target non-zero ratio of {\ShflBW} pruning (denoted as $\alpha$). In practice, we find a reduced sparsity, $\beta = 2\alpha$, produces better final results. This unstructured pruning step  generate{s} a binary mask as matrix \texttt{(c)}{,} indicating the positions of remained weights. Next, we invoke the K-Means algorithm to cluster the rows in the binary mask into groups with a fixed size, $V$. In this example, each group contains $V=2$ rows. Next, we permute the rows of the original weight matrix to put rows in the same group together and get matrix \texttt{(d)}. After reordering the rows, we apply a normal vector-wise pruning to the target non-zero ration $\alpha$ and get the matrix in \texttt{(e)}. As the final step, we {reverse} the row permutation applied to matrix \texttt{(d)}, recovering the original row ordering of weights.

\section{Evaluation} \label{sec:eval}
\begin{figure}[t]
    \includegraphics[width=\linewidth]{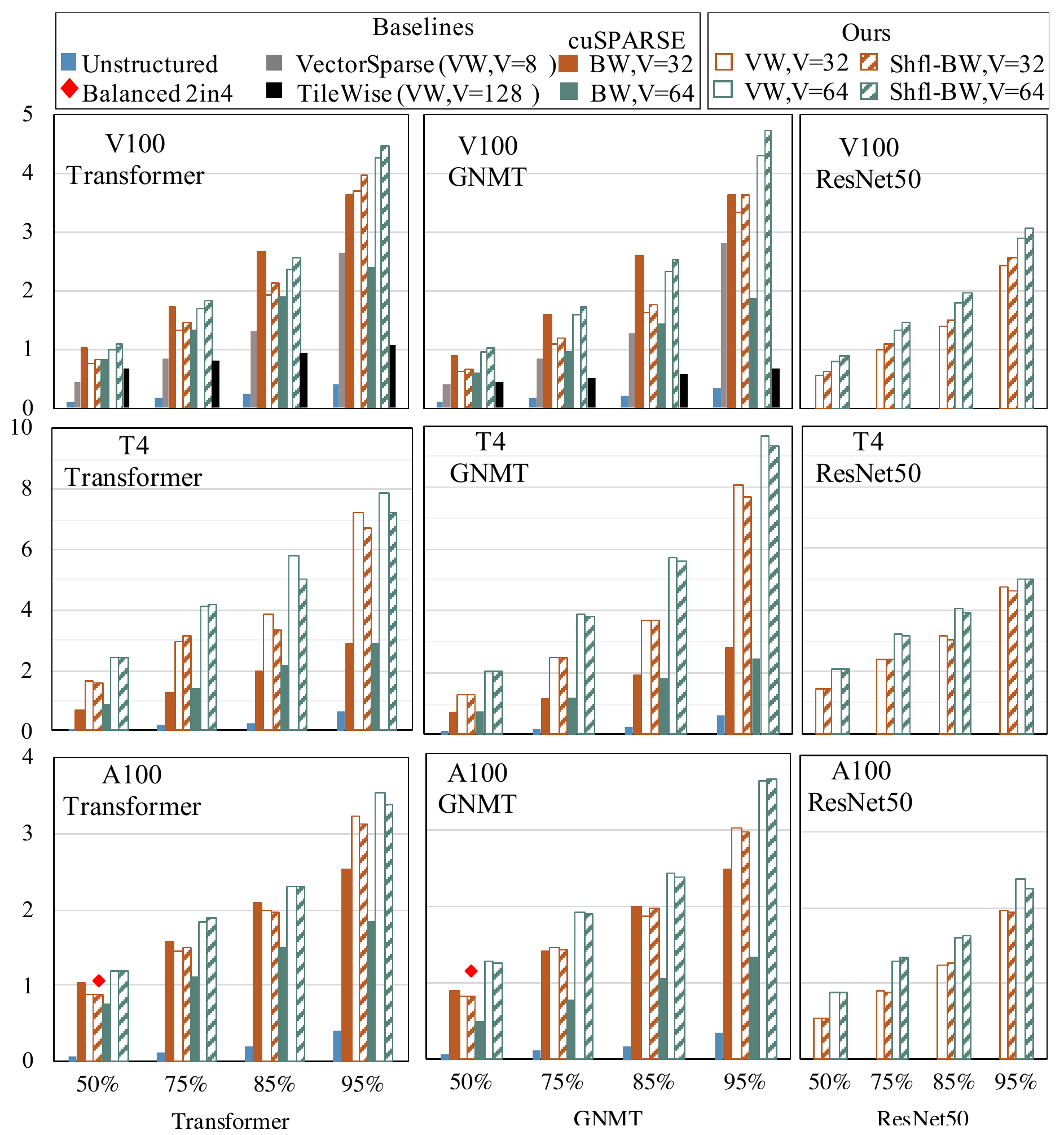}
    \caption{Speedup against the dense baseline on three GPUs and three models.}\label{fig:eval:kernel}
\end{figure}
\subsection{Methodology}
% We evaluate our solution  include results in pruning accuracy and model acceleration. The first part is overall results, where we compare us against two most relevant papers in terms of speedup-accuracy trade-off curves. The second and third part evaluate in detail the kernel speedup and model pruning results against the two prior papers and other kernel-only or pruning-only baselines.

\textbf{Workload}: We {experiment} on three DNN models: Transformer~\cite{vaswani2017attention} and GNMT~\cite{wu2016google} on the WMT translation task, and ResNet50~\cite{he2016deep} for ImageNet classification. %Transformer are mainly composed of linear layers {for matrix-matrix multiplication operations} . GNMT are composed of LSTM layers which involve many {matrix vector operations}. ResNet50 is composed of 2D convolutions. 
We only calculate the speedup to the linear and 2D convolution layers in the following results. %We implement the weight-sparse convolution with Implicit-GEMM algorithm~\cite{cudnn} with similar techniques as introduced in Section.~\ref{sec:kernel}. %The model inference also involves layers that cannot be accelerated by weight pruning, but GEMM and convolution take up a major percentage. For example, GEMM takes up ~\cite{Guo2020}. 
When reporting model kernel speedup, we use the shapes in real model.

\textbf{Platform}: We test on NVIDIA V100, T4 and A100 GPUs, {which are}  the latest three architectures equipped with tensor-cores.

\textbf{GPU kernel baselines}: We compare {\ShflBW} with dense, unstructured, block-wise (\textbf{BW}) and vector-wise (\textbf{VW}) sparse patterns. We use cuBLAS~\cite{cublas} and cuDNN~\cite{cudnn} as the dense baseline for GEMM and 2D convolution respectively.  {`Unstructured'} baseline comes from Sputnik~\cite{Gale2020}. `Balanced` sparse kernels is supported by A100 at 50\% sparsity only, and we evaluate the SpMM kernel in NVIDIA's library cuSPARSELt~\cite{cusparselt}.   %, a library optimized for DNN sparsity characteristics \hr{;} 
Block-wise baseline comes from NVIDIA cuSPARSE~\cite{cusparse} library. cuSPARSE does not provide vector-wise kernels, so we implement our own, and in addition we compare with two prior papers on vector-wise kernels: Tilewise~\cite{Guo2020}, which uses CUDA multi-streams, and VectorSparse~\cite{Chen2021}, which is tuned for fine-grained vector-wise SpMM ($V\leq8$) on tensor-cores. 

\textbf{Pruning settings}: We train the GNMT model based on the ADMM~\cite{zhang2018systematic} method, which improves final accuracy through changing the weight distribution before the pruning step. For Transformer and ResNet50, we apply a newly published workflow named Grow-and-Prune{~\cite{ma2021effective}}, which improves the quality of pruned model through multiple round of growing and pruning.
\subsection{Kernel Speedup}
\textbf{Overall results}: Figure~\ref{fig:eval:kernel} shows the speedup over dense kernels for three DNN models, correlated to the sparsity. The baselines all lack implementation for convolution. Tilewise~\cite{Guo2020} and VectorSparse~\cite{Chen2021} both need to be compiled on V100 only. $V$ refers to the block (or vector) size in block-wise, vector-wise and {\ShflBW}.The throughput of {\ShflBW} sparse kernels increase along with sparsity and $V$. \guyue{For example, on T4 GPU for Transformer GEMM layers, {\ShflBW} gives speedup of  $3.17/4.18\times$ under $V=32/64$, 75\% sparsity, and $3.35/5.04\times$ under $V=32/64$, 85\% sparsity. }
At 75\% sparsity, We accelerate the GEMM layers of Transformer~\cite{vaswani2017attention} by up to \textbf{1.81$\times$, 4.18$\times$, 1.90$\times$} on V100, T4 and A100 GPUs respectively. The speedup on T4 is higher than V100 and A100, because T4 has lower ratio of computation capability to bandwidth, posing smaller demand on data reuse. 

\textbf{Comparing sparsity patterns}: Unstructured sparsity cannot exceed the dense baseline at even 95\% sparsity, because it does not use tensor-cores. {\ShflBW} performs comparable to block-wise, while the cuSPARSE block-wise SpMM shows unstable performance across GPUs and block sizes. For example, {\ShflBW} is in average 2.88$\times$ cusparse block-wise on T4 GPU at V=64, but only 0.83$\times$ on V100 at $V=32$. Compared to our vector-wise implementation, {\ShflBW} is in average $0.97-1.02\times$ faster, showing that \textbf{row shuffling involves negligible overhead with our reordered write-back technique}. 
Finally, our evaluation shows that on A100, balanced sparsity gives only $1.07\times,1.16\times$ speedup on A100 for Transformer and GNMT, which is  less {performant} than {\ShflBW} at $V=64$ at 50\% sparsity. 

\textbf{Comparing kernel designs}: Our vector-wise and {\ShflBW} implementation consistently outperform two vector-wise prior papers. Due to the overhead when {the} number of streams grows, their multi-stream approach cannot exceed dense baseline under real weight shapes\f{ without the neuron pruning applied described in the paper~\cite{Guo2020}}
\comments{\f{what is neuron pruning?}}. VectorSparse~\cite{Chen2021} is less {performant} than ours because their small vector size ($V=8$) limits data reuse.
\begin{table}[t]
\begin{tabular}{cc|ccc}
\hline
\small
                                           &                 & \begin{tabular}[c]{@{}c@{}}Transformer\\ ({BLEU})\end{tabular} & \begin{tabular}[c]{@{}c@{}}GNMT\\ ({BLEU})\end{tabular} & \begin{tabular}[c]{@{}c@{}}ResNet50\\ ({Top-1 Acc.}\%)\end{tabular} \\ \hline
\multicolumn{1}{c|}{\multirow{4}{*}{80\%}} & BW, V=32         & 26.92                                                              & 13.83                                                       & 75.51                                                     \\
\multicolumn{1}{c|}{}                      & VW, V=32         & 26.97                                                              & 23.1                                                        & 75.77                                                     \\
\multicolumn{1}{c|}{}                      & Shfl-BW, V=32    & 27.23                                                              & \textbf{24.09}                                              & \textbf{76.02}                                            \\
\multicolumn{1}{c|}{}                      & Shfl-BW, V=64    & \textbf{27.25}                                                     & 23.88                                                       & 75.94                                                     \\ \hline
\multicolumn{1}{c|}{\multirow{3}{*}{90\%}} & VW, V=32         & 26.1                                                               & 21.27                                                       & 72.69                                                     \\
\multicolumn{1}{c|}{}                      & Shfl-BW, V=32    & 26.29                                                              & \textbf{23.3}                                               & \textbf{73.83}                                            \\
\multicolumn{1}{c|}{}                      & Shfl-BW, V=64    & \textbf{26.53}                                                     & 22.68                                                       & 73.09                                                     \\ \hline                                   
\end{tabular}
\caption{BLEU score and accuracy of pruned models with different sparse patterns.}
\label{tab:accuracy}
\end{table}

\subsection{Model Accuracy}

As shown in Figure~\ref{tab:accuracy}, at 80\% sparsity, {\ShflBW} consistently improves the model quality over vector-wise and block-wise. For example on Transformer, {\ShflBW} with $V=32$ shows 0.37 drop of BLEU score from the unstructured, but vector-wise at $V=32$ involves 0.63. In most cases, the quality of $V=64$ pruning with shuffling gives better results than a smaller block size but without shuffling.

\section{Conclusion} \label{sec:concl}
Although tensor-core\f{s} significantly {increase} the throughput of GPUs, it poses challenges on exploiting sparsity for accelerating DNN inference. This paper proposes a new sparsity pattern, {\ShflBW}, to strike a better trade-off between {\flexibility} and {\performability}. \guyue{We propose GPU sparse kernel designs to exploit tensor-core acceleration.
% \guyue{(e.g. Transformer at 75\% sparsity by 4.18$\times$ on T4 GPU)}
{\ShflBW} is comparably efficient but much more flexible than block-wise sparsity. We expect our tensor-core aware pruning methodology to {be applicable to a variety of} applications, given the emerging trend of adding tensor-core-like units to DNN processors. }

\section*{Acknowledgements}
This work was supported by Alibaba Group through Alibaba Research Intern Program.

\tiny
\bibliographystyle{unsrt}
\bibliography{main}

\end{document}